\documentclass[runningheads]{svmult}

\usepackage{makeidx}   
\usepackage{graphicx}  
\usepackage{subeqnar}  
\usepackage{multicol}  
\usepackage{physprbb}  
\makeindex             

\begin{document}
\title*{The dusty type IIn Supernova 1998S}
\toctitle{The dusty type IIn Supernova 1998S}
%
%
\titlerunning{The dusty type IIn Supernova 1998S}
%
\author{Peter Meikle\inst{1}
\and Alexandra Fassia\inst{1}
\and Thomas R. Geballe\inst{2}
\and Peter Lundqvist\inst{3}
\and Nikolai Chugai\inst{4}
\and Duncan Farrah\inst{1}
\and Jesper Sollerman\inst{3}}
\authorrunning{Peter Meikle et al.}
%
%
\institute{Blackett Laboratory, Imperial College, Prince Consort Road,
London SW7 2BW, UK 
\and Gemini Observatory, 670 N. A'ohoku Place, Hilo, HI 96720, USA
\and Stockholm Observatory, AlbaNova, Dept. of Astronomy, Stockholm SE
106 91, Sweden
\and Insitute of Astronomy, RAS, Pyatnitskaya 48, 109017, Russia
}

\maketitle              

\begin{abstract}
The type~IIn SN~1998S is one of the most remarkable core-collapse
supernovae ever observed.  It underwent a complex interaction with a
substantial circumstellar medium, resulting in radiation at
wavelengths from radio to X-rays.  IR and optical observations have
revealed a wide variety of broad and narrow emission lines.
Examination of the SN/CSM interaction and of the ejecta spectra has
allowed us to deduce that the supernova probably arose from a massive,
RSG progenitor having a large ($>$3200~AU radius), dusty circumstellar
disk.  SN~1998S also developed one of the strongest, most persistent
infrared excesses ever seen in a supernova.  IR/optical monitoring of
SN~1998S has been carried out to nearly 1200~days post-explosion.
This includes coverage to wavelengths as long as 4.7~$\mu$m, making
SN~1998S only the second supernova (after SN~1987A) to be observed in
this spectral region. Fading of the central and redshifted components
of the late-time H~I and He~I line profiles suggests strongly that
dust condensed in the ejecta.  However, it is less clear whether the
strong late-time IR emission arose from this dust, or from an IR echo
in the dusty CSM.  One interesting possibility is that dust condensed
in the cool dense shell between the outer and reverse shocks, thus
simultaneously producing both the line obscuration and the IR
emission.\footnote{Invited talk given at ESO/MPA/MPE Workshop {\it
From Twilight to Highlight: The Physics of Supernovae}, Garching,
Germany, 29-31 July 2002. To appear in the Proceedings,
eds. W. Hillebrandt \& B. Leibundgut (pub. Springer-Verlag series
``ESO Astrophysics Symposia''.)}
\end{abstract}

\section{Introduction}
One of the challenges of supernova research is to obtain evidence
about the nature and environment of the progenitor.  For type IIn
supernovae, the progenitors must have undergone one or more mass-loss
phases before explosion.  By using the `illumination' of the resulting
circumstellar medium (CSM) by the supernova explosion we can get clues
about the progenitor, even at distances of tens of Mpc. \\

More than 30 years ago, it was suggested [1--3] that supernovae could
be important sources of interstellar dust.  More recent work [4--6]
still invokes core-collapse SNe as significant dust contributors.
However, the number of supernovae in which dust has been detected is
relatively small and, prior to SN~1998S, only for SN~1987A had dust
condensation been convincingly demonstrated [7--14]. \\

An opportunity to address both the progenitor problem and the question
of supernovae as dust sources has been provided by the occurrence of
the type~IIn SN 1998S.  This has become the most intensively-studied
type IIn event [15--30].

\section{Early-time behaviour}
\subsection{Early-time optical spectral evolution}
The earliest optical spectra of SN~1998S show a blue continuum with
emission features superimposed. A rough blackbody fit yields
T$\sim$25,000~K, but with a blue excess [18, 24].  The emission lines
are identified with H~I (Balmer series), He~I, He~II, C~III and N~III.
The high-ionisation carbon and nitrogen lines are also commonly
observed in Wolf-Rayet stars [18]. The emission lines have a broad
base ({\it e.g.}  H$\alpha$ FWZI$\sim$20,000 km/s), but a narrow
`peaked' unresolved centre.  The lines are symmetrical about the local
standard of rest.  This is quite surprising since at such an early
phase most of the receding part of the supernova should be occulted by
the photosphere.  In fact, Chugai [23] has shown that this constitutes
some of the earliest evidence of a strong ejecta-CSM wind
interaction. The broadening results from Thomson scattering in a
radiatively-accelerated CSM wind lying immediately above an opaque,
relatively cool dense shell (CDS) at the ejecta-wind interface
[31,32].  The blue excess can also be attributed to the CDS, since the
significant optical depth can yield an increase in continuum
absorptive opacity with wavelength, due to both bound-free ({\it i.e.}
the Paschen Continuum) and free-free processes [18].\\

By about 2~weeks after the explosion, the emission lines had
essentially disappeared. (Following Fassia et al. [24], we adopt JD
2450875.2 as zero epoch. This was probably a few days post-explosion.)
This disappearance is attributed to the dense inner-CSM being overrun
by the ejecta.  Nevertheless, the CDS remained optically-thick in the
Paschen continuum until around 40-50 days, and this accounts for the
lack of strong broad lines from the ejecta during the $\sim$2--6 weeks
era. However, during this time we can see weak unresolved lines
superimposed on the continuum.  This is due to the flash-ionisation of
the undisturbed wind (see below).  After $\sim$40--50~days, broad,
square-shaped lines in H${\alpha}$ and the Ca~II triplet formed.  Such
line profiles are characteristic of emission from a shocked ejecta/CSM
shell.

\subsection{Early-time infrared spectral evolution}
SN~1998S is unique in that it allowed the first-ever good IR
spectroscopic coverage of a type~IIn event [17,24].  In the $J$-band,
we see Paschen $\beta$, Paschen $\gamma$ and He~I 1.083~$\mu$m lines.
Their evolution was similar to that seen in the optical.  At the
earliest times broad-based, peaked profiles were present. These faded
by day~17, being replaced by broad, square-shaped profiles by day~44.
Between days $\sim$10 and $\sim$60 a strong, unresolved He~I
1.083~$\mu$m CSM line was superimposed on the ejecta/CSM broad lines.\\

By day~44, the $HK$-band was dominated by Paschen~$\alpha$.  By
day~108, first-overtone CO emission was clearly present in the
$K$-band.  The presence of CO in core-collapse supernovae is
increasingly regarded as ubiquitous.  In all cases where $K$-band
observations have been carried out in the period 3-6 months
post-explosion, CO has been detected. There are now seven known cases:
87A (IIpec) [13,33-35], 95ad (IIP) [36], 98S (IIn) [17,24], 99dl (IIP)
[37], 99em (IIP) [37,38], 99gi (IIP) [38], 00ew (Ic) [38].  Modelling
of the SN~1998S spectra suggests a CO velocity of $\sim$2000 km/s
[17,24].  From this, Fassia et al. [24] deduced a core mass of
4~M$_{\odot}$ implying a massive progenitor. The actual mass of CO
derived was 10$^{-3}$~M$_{\odot}$.  The low-excitation
rotation-vibration states of CO mean that it is a powerful coolant.
Its presence is suspected to be a necessary condition for dust
condensation to occur in the ejecta. \\

On day~130, Fassia et al [24] succeeded in measuring the IR flux out
to a wavelength of 3.8~$\mu$m ($L'$-band).  This revealed a remarkable
IR excess of $K-L'=+2.5$.  The most plausible interpretation of this
is emission from warm dust. But where was the dust located?  For dust
condensing in the ejecta to produce such a large, early $L'$ flux the
lowest possible velocity of the dust-forming region would be
11,000~km/s, and that includes the assumption that the temperature is
close to the dust evaporation temperature of $\sim$1500~K. Such high
velocities were seen only in the extreme outer zones of the H/He
envelope. No metals were seen at such high velocities.  Fassia et
al. [24] concluded that the IR excess at this epoch cannot, therefore,
have been due to grain condensation in the ejecta.  It must instead
have been produced by an IR echo of the maximum-light luminosity from
pre-existing dust in the CSM.
 
\subsection{Narrow lines}
Of particular interest are the high-resolution echelle spectra of SN
1998S obtained at the WHT by Bowen et al. [15] and Fassia et al. [24]
on days~17 and 36.  These observations succeeded in resolving the
narrowest CSM lines.  From forbidden lines such as [OIII]~5007~\AA, an
undisturbed CSM velocity of about 40~km/s is obtained, which is
characteristic of an RSG wind.  Fassia et al. [24] also deduced a
centre of mass redshift velocity of +847~km/s.  Between day~17 and
day~36 the [OIII]~5007~\AA\ profile changed from having a red deficit,
to being quite symmetrical about the SN centre of mass. This is
attributed to the effect of the finite light travel time across the
CSM [24].  As the initial ionising flash from the supernova propagated
across the CSM, it took longer for the resulting nebular emission to
reach us from the far side.  This allowed confirmation that the CSM
really was expanding.  Making some simple assumptions, the echo
geometry indicates that the CSM extended to at least $\sim$2100~AU.\\

The narrow [OIII]~5007~\AA\ line persisted for at least a year [28].
Assuming a maximum ejecta velocity of 10,000 km/s [18,24], it can be
deduced that the unshocked CSM must have extended to at least 2000~AU,
which is consistent with the lower limit derived from the echo
interpretation.  In fact, later observations have shown that the CSM
extended to at least 3200~AU (see below).  From the intensity ratio of
[OIII] (4959~\AA\ + 5007~\AA) to [OIII] 4363~\AA, Fassia et al.  [24]
infer a wind density of at least $1.5\times10^6$~cm$^{-3}$, implying a
CSM mass exceeding 0.005~M$_{\odot}$, and a mass-loss rate exceeding
around $2\times10^{-5}$~M$_{\odot}$/yr.  This is consistent with the
radio/X-ray estimate of around 10$^{-4}$~M$_{\odot}$/yr [30].\\

The behaviour of the allowed H~I, He~I CSM lines was more complex.
Not only did they exhibit asymmetric P~Cygni profiles, but there were
clearly two velocity components.  The slower component is attributed
to the same origin as the forbidden lines {\it viz.} the
photo-ionised, unaccelerated CSM.  The profile of this component was
probably a combination of emission from the recombination cascade
together with a classical P~Cygni line due to scattering from the
populated excited levels (resulting from the recombination cascade).
The broad absorption component is more difficult to explain.  It
extends to a velocity of around 350~km/s which is too fast for a red
supergiant wind.  It may be that, as in the case of SN~1987A, the
SN~1998S progenitor went through a fast-wind phase prior to explosion
[24].  An alternative explanation is that CSM close to the supernova
was accelerated by photospheric photons, or by relativistic particles
from the ejecta/CSM shock [26]. Another possibility is that the faster
component arose in shocked clumps within the CSM wind [26]. 

\subsection{Bolometric Light Curve}
SN~1998S was exceptionally luminous, reaching a de-reddened M$_B =
-19.6$ [16]. This is around $\times$10 the typical luminosity of a
type~II SN.  The excellent coverage achieved in the optical and IR
allowed Fassia et al. [16] to examine the bolometric light curve.
Both blackbody and UVOIR fits indicate that the total energy radiated
in the first 40~days exceeded $10^{50}$ ergs, which is again
$\times$10 the typical value for type~II SNe.  Between days~90 and
130, the bolometric light curve is well-reproduced by the radioactive
decay luminosity of 0.15~M$_{\odot}$ $^{56}$Ni.  However, by this era
the ejecta/CSM shock energy must also have contributed a minor
contribution.

\subsection{Polarisation}
Spectropolarimetry by Leonard et al. [18] and Wang et al. [25]
indicate asymmetry in the material responsible for the observed
radiation.  Leonard et al. favour a highly flattened CSM, with
possibly some asymmetry in the ejecta.  In contrast, Wang et
al. favour ejecta asymmetry as the main cause of the polarisation.

\section{SN~1998S at late times}
Most of the observed features described in the previous section can be
attributed to the interaction of the supernova with a pre-existing,
dusty, possibly flattened CSM.  SN~1998S remained observable from
X-rays to radio for over 3 years [30,39].  This persistence was due to
the ongoing conversion of the SN kinetic energy to radiation via the
ejecta/shock interaction.  To obtain further insights into this
phenomenon, regular observations continued during this phase.  In this
section, I shall consider two aspects of this {\it viz.} the IR
emission and the nature and evolution of the line profiles.

\subsection{The infrared spectral energy distribution at late times}
Infrared monitoring of SN~1998S continued at UKIRT up to day~1191
[39].  Observations were extended as far as the $M$-band
(4.7~$\mu$m). Other than SN~1987A, this is the only time that such
longwave IR radiation has been detected from a supernova.  The IR
excess persisted throughout this period.  Between days 326 and 819,
plausible blackbody fits to the de-reddened $HKL'M$ photometry
(1.6--4.7~$\mu$m) are obtained.  (We exclude the $J$-band to avoid
contaminating the analysis with the very strong He~I 1.083~$\mu$m
emission.)  The derived temperature and velocity declined from around
1400~K and 4000~km/s on day~326, to 930~K and 2000~km/s on day~819.
However, for the latest photometry (days~1042 and 1191) it was not
possible to achieve a single-temperature fit.  On day~328, IR
spectroscopy to 2.5~$\mu$m was acquired at UKIRT.  This revealed that,
while there was a small contribution due to Paschen~$\alpha$, the IR
excess was due primarily to a smooth continuum rising to longer
wavelengths.  We conclude that the late-time IR excess was due to
thermal emission from warm dust.\\

We can now pose two, possibly connected, questions: what powered the
IR emission from the dust, and where was the dust located?  The total
energy emitted by the dust in the 1.6--4.7~$\mu$m region between days
300 and 1200 was about 10$^{49}$~ergs.  This is a factor of $\times$10
more than could be supplied by the decay of the daughter products of
0.15~M$_{\odot}$ $^{56}$Ni over the same period.  We can therefore
immediately rule out radioactivity as the source of the IR energy.
There are two other possible energy sources. We know that the energy
of the early light curve amounted to $\sim$10$^{50}$~ergs.  Thus, one
possibility is that 10\% of this was channeled into the IR emission
via an IR echo from CSM dust. On the other hand, it is likely that of
order 10$^{51}$~ergs was stored in the kinetic energy of the
ejecta. It would take only 1\% of this to account for the IR emission,
through the heating of either pre-existing (CSM) or newly-condensed
(ejecta) dust.  At 130~days, the huge $L'$-band flux argues strongly
in favour of emission from pre-existing dust {\it i.e.} an IR echo.
However, such an argument is less convincing at the later times being
considered here.  The blackbody fits produce velocities of
4000--2000~km/s which could, just conceivably, have arisen from dust
condensation in the ejecta. To try to distinguish between the IR echo
and dust condensation scenarios, we now examine the line profiles at
late times.

\subsection{H${\alpha}$ and He~I 1.083~$\mu$m profiles}
The H$\alpha$ profile of SN~1998S changed quite dramatically between
day~97 and day~1086.  On day~97 the profile had the form of a broad,
steep-sided, fairly symmetrical line spanning $\pm$7000~km/s across
the base [24].  This appearance persisted to at least day~140 [18].
However, by the time the supernova was recovered in the second season,
the shape was remarkably different. The day~240 (relative to our
adopted zero epoch) spectrum of Gerardy et al. [17] shows that the
profile had developed a triple-peak structure, comprising a central
peak close to the rest-frame velocity, and two outlying peaks at,
respectively, $\pm$4500~km/s.  Gerardy et al. suggest that the
outermost peaks could have been produced by an emission zone having a
ring or disk structure seen nearly edge-on, and resulting from the SN
shockwave collision with the disk/ring.  Following Chugai \& Danziger
[40] they also suggest that the central peak might have been due to
shocked wind clouds.  Spectra obtained on days~276 [17] and 288 [39]
show a remarkable fading of the central and redshifted peaks with
respect to the blueshifted peak.  Subsequent spectra to day~640 show
a continuation of this trend [ref. 39, R. Fesen private communication,
A. Filippenko private communication]. In addition, the strong blue
peak shifted in velocity to --3500~km/s, presumably due to a slowing
of the shock as it encountered an increasing mass of CSM. Finally,
when the supernova was recovered in the 4th season on day~1086, it was
found that the profile had undergone another dramatic change [39].
While the blue-shifted peak at about --3500~km/s persisted, the
central peak had grown in relative strength to about twice the height
of the blue peak.  An explanation for this latest behaviour is still
being investigated.  IR ($J$-band) spectra were obtained at UKIRT
between days~225 and 1185 [39]. IR spectra were also acquired by
Gerardy et al. on days 276 and 370 [17].  The form and evolution of
the strong He~I 1.083~$\mu$m profile was very similar to that of
H$\alpha$.  The H$\alpha$ and He~I 1.083~$\mu$m lines persisted to
$>$1100~days by which time their extreme blue limbs were still at a
velocity of $\sim$5000~km/s.  This indicates that the CSM must have
extended to $>$3200~AU.

\subsection{Source of the IR emission and location of the dust}
The relatively sudden fading of the central and redshifted components
of the H~I, He~I lines immediately suggests dust condensation in the
ejecta.  This could have caused obscuration of the central and
receding regions.  A similar effect was observed in the ejecta line
profiles of SN~1987A [12,41] and, more recently, in the type IIP
SN~1999em [42].  The presence of CO emission from SN~1998S as early as
130~days lends credence to the dust condensation scenario.  Moreover,
the effect is comparably strong at 0.66~$\mu$m and 1.08~$\mu$m,
suggesting that the dust quickly became optically thick.  It is
difficult to see how pre-existing dust in the CSM could have produced
such an effect.  Indeed, pre-existing dust would probably have been
evaporated by the initial flash out to a radius of at least several
thousand~AU [43].  On the other hand, there may actually have been two
dust zones present.  In this more complicated scenario, the IR
emission would be due to an IR echo of the SN peak luminosity from
pre-existing dust in the CSM. The line-profile obscuration, however,
would be due to possibly cooler dust condensing in the ejecta.\\

We note an interesting coincidence. Throughout the second year, the
magnitude and evolution of the velocities of the blue-shifted H~I,
He~I peaks were similar to those derived from the blackbody fits to
the IR fluxes.  It has been recognised for many years ({\it e.g.}
ref. 31) that the interaction of the supernova ejecta with a dense CSM
will produce outer and reverse shocks.  When radiative cooling is
important at the reverse shock front, the gas undergoes a thermal
instability, cooling to $\sim$10,000~K, thus forming a dense,
relatively cool zone - the `cool dense shell' or CDS.  Line emission
from low-ionisation species in the CDS will be produced [32].  We
believe that this emission was responsible for the blueshifted and
redshifted peaks of the H$\alpha$ and He~I line profiles.  An exciting
possibility, which still requires further study, is that dust may have
formed in the CDS at the ejecta/wind interface.  If cooling in the
outer layer of the CDS, shielded from the reverse shock X-ray/UV
radiation, brought the temperature to below the condensation
temperature, dust could have formed and survived there.  In a similar
process, suggested by Usov [44], dust may form in the colliding winds
of Wolf-Rayet stars.  Rayleigh-Taylor or convective instabilities [32]
might have produced opaque clumps of dust, totally obscuring the
central and receding parts of the supernova, while at the same time
allowing some of the line radiation to escape from the approaching
component of the CDS.  Thus, this scenario can simultaneously account
for the strong IR flux, the obscuration effect and the velocity
coincidence with the line profiles.

\section{Summary}
The detailed study of the type IIn SN~1998S indicates that it probably
arose from a massive, RSG progenitor having a large ($>$3000~AU),
dusty circumstellar disk.  The excess IR emission at early times was
due to an IR echo from this disk. At late times the origin of the
strong IR emission is less clear. It may be that a `double-dust'
scenario applies where the line obscuration was due to dust
condensation in the ejecta, while the IR emission arose from an IR
echo from the dusty CSM.  Alternatively, dust condensation in the cool
dense shell may account for both the line obscuration and the IR
emission. More detailed modelling of the data will be required in
order to test this `single-dust' scenario.

%

\end{document}